\documentclass[aps,prd,preprintnumbers,reprint,onecolumn,groupedaddress,nofootinbib]{revtex4-2}

\usepackage{graphicx}  % needed for figures
\usepackage{dcolumn}   % needed for some tables
\usepackage{bm}        % for math
\usepackage{bbm}        % for math
\usepackage{amssymb}   % for math
\usepackage{amsmath}
\usepackage{slashed}
\usepackage{mathtools}
\usepackage[colorlinks=true,linkcolor=red]{hyperref}
\usepackage{cleveref}
\usepackage[caption=false]{subfig}
\usepackage[export]{adjustbox}

\usepackage{eso-pic}
\newcommand{\reportnum}[2]{
    \AddToShipoutPictureBG*{%
        \AtPageUpperLeft{  
            \hspace{0.75\paperwidth}
            \raisebox{#1\baselineskip}{
                \makebox[0pt][l]{\textnormal{#2}}
            }
        }
        }
}
\reportnum{-2.5}{SAGEX-21-22-E}

\allowdisplaybreaks

\newcommand{\cA}{\mathcal{A}}

\newcommand{\cM}{\mathcal{M}}

\newcommand{\cS}{\mathcal{S}}
\newcommand{\cT}{\mathcal{T}}

\begin{document}

% the following line is for submission, including submission to the arXiv!!
%\hspace{5.2in} \mbox{Fermilab-Pub-04/xxx-E}

\title{Exponentiation of the leading eikonal with spin}
%\input author_list.tex       % D0 authors (remove the first 3 lines
                             % of this file prior to submission, they
			     % contain a time stamp for the authorlist)
\author{Kays Haddad}                             % (includes institutions and visitors)

\email[]{kays.haddad@nbi.ku.dk}

\affiliation{Niels Bohr International Academy and Discovery Center, Niels Bohr Institute,
University of Copenhagen, Blegdamsvej 17, DK-2100 Copenhagen, Denmark}

\date{\today}

\begin{abstract}
We initiate a study into the eikonal exponentiation of the amplitude in impact-parameter space when spinning particles are involved in the scattering.
Considering the gravitational scattering of two spin-1/2 particles, we demonstrate that the leading eikonal exhibits exponentiation up to $\mathcal{O}(G^{2})$ in the limit where the spacetime dimension $D\rightarrow4$.
We find this to hold for general spin orientations.
The exponentiation of the leading eikonal including spin is understood through the unitarity properties at leading order in $\hbar$ of momentum-space amplitudes, allowing the extension of our arguments to arbitrary-spin scattering.
\end{abstract}

\pacs{}
\maketitle

%\AddToShipoutPictureBG*{%
%  \AtPageUpperLeft{%
%    \hspace{\paperwidth}%
%    \raisebox{-1.75\baselineskip}{%
%      \makebox[0pt][r]{SAGEX-21-22-E\qquad}
%}}}%

\section{\label{sec:intro}Introduction}

The application of scattering amplitude and quantum-field theoretic techniques to classical systems has expanded rapidly in recent years, largely motivated by describing the inspiral phase of compact binary coalescence.
The dynamics of spinless inspiraling black holes have been fully understood up to the third post-Minkowskian (PM) order \cite{Cheung:2018wkq,Bern:2019nnu,Bern:2019crd,Cheung:2020gyp,Herrmann:2021lqe,Herrmann:2021tct,DiVecchia:2020ymx,DiVecchia:2021ndb,DiVecchia:2021bdo,Bjerrum-Bohr:2021din}, and results for conservative effects at 4PM have already emerged \cite{Bern:2021dqo} (see also refs.~\cite{Kalin:2020fhe,Dlapa:2021npj} for analogous results obtained in the purely classical setting of ref.~\cite{Kalin:2020mvi}).

Alongside this progress, the understanding of the relation between classical and quantum spin has itself seen substantial development \cite{Chung:2018kqs,Guevara:2018wpp,Maybee:2019jus,Guevara:2019fsj,Chung:2019duq,Arkani-Hamed:2019ymq,Aoude:2020onz,Bern:2020buy,Guevara:2020xjx,Aoude:2021oqj}.
Scattering dynamics are understood for arbitrary spin at 1PM order \cite{Guevara:2018wpp,Chung:2019duq,Guevara:2019fsj,Aoude:2020onz,Aoude:2021oqj}, while progress at higher PM orders presently lays at 2PM and quadratic order in spin \cite{Damgaard:2019lfh,Bern:2020buy,Kosmopoulos:2021zoq}.
Results at the 2PM order quartic in the spin of each scattering particle exist in the aligned-spin setup \cite{Guevara:2018wpp}.
Progress in this direction is restricted by the lack of a unique gravitational Compton amplitude for a massive particle with spin $s>5/2$ \cite{Arkani-Hamed:2017jhn,Chung:2018kqs,Aoude:2020onz}, with the $s=5/2$ Compton amplitude being fixed only recently \cite{Chiodaroli:2021eug}.

In addition to the description of point particles, tidal effects -- both with and without spin -- are also describable using a quantum-field-theoretic approach \cite{Cheung:2020sdj,Haddad:2020que,Bern:2020uwk,AccettulliHuber:2020dal,Aoude:2020ygw}.

Phenomenological predictions have been made accessible to amplitudes-based techniques thanks to various formalisms bridging the gap between quantum field theory and classical physics \cite{Cheung:2018wkq,Kosower:2018adc,Maybee:2019jus,Cristofoli:2019neg,Kalin:2019rwq,Bjerrum-Bohr:2019kec,Cristofoli:2020uzm,Bern:2020buy,Cristofoli:2020hnk,Mogull:2020sak,Jakobsen:2021lvp,Cristofoli:2021vyo,Bautista:2021wfy,Aoude:2021oqj,delaCruz:2021gjp}.
One such path to classical observables makes use of the eikonal phase, related to the classical portion of the scattering amplitude in impact-parameter space \cite{Levy:1969cr,Amati:1987uf,DiVecchia:2019myk,DiVecchia:2020ymx,DiVecchia:2021ndb,DiVecchia:2021bdo}.
The exponential of the eikonal phase relates the scattering amplitude in impact-parameter space to several classical observables at all orders in Newton's constant $G$, such as the scattering angle \cite{Amati:1987uf, DiVecchia:2021bdo}, the linear impulse, and the spin kick \cite{Bern:2020buy}.\footnote{The relationship between classical observables and the exponentiated spinning eikonal phase in ref.~\cite{Bern:2020buy} is conjectural at $\mathcal{O}(G^{3})$ and above.}

Despite the ubiquitous application of the eikonal phase with spin to the derivation of classical observables from scattering amplitudes \cite{Guevara:2018wpp,Guevara:2019fsj,Arkani-Hamed:2019ymq,Bern:2020buy,Aoude:2020ygw,Aoude:2021oqj},
the exponentiation properties of the gravitational impact-parameter space amplitude in this setup have not been investigated.\footnote{Studies of the eikonal including spin exist in non-gravitational contexts; see e.g. refs.~\cite{Banerjee:1977na,Mcneil:1979aa,Waxman:1981ve}. We further expand on these by considering arbitrary spin in \Cref{sec:Unitarity}.}
In this note, we take first steps towards filling this gap.
Computing the scattering of two spin-1/2 particles up to $\mathcal{O}(G^{2})$, we do indeed find a relation in impact-parameter space between the square of the leading eikonal phase (the eikonal phase at $\mathcal{O}(G)$) and the super-classical divergences of the one-loop amplitude in the limit where the spacetime dimension $D\rightarrow4$.
This relation is suggestive of the exponentiation of the eikonal phase even in the presence of massive spinning matter.
We understand this exponentiation through the unitarity properties of the amplitude, which are simplified for spinning amplitudes at leading order in $\hbar$.

We begin in \Cref{sec:Eikonal} with an introduction to the eikonal phase.
In \Cref{sec:LeadingEikonal} we calculate the leading eikonal for the scattering of two spin-1/2 particles, and relate the square of the eikonal to the super-classical divergences of the one-loop spin-1/2 $\times$ spin-1/2 amplitude.
In \Cref{sec:Unitarity} we analyze the leading-in-$\hbar$ unitarity relations of amplitudes containing first spin-1/2, then spin-$s$ particles, and make a connection to the exponentiation of the leading eikonal phase.
We conclude in \Cref{sec:Conclusion}.

\section{The eikonal phase}\label{sec:Eikonal}

We begin with a brief introduction to the eikonal phase and its relation to the impact-parameter space amplitude.
For more details, see e.g. refs.~\cite{Levy:1969cr,KoemansCollado:2019ggb,DiVecchia:2019myk,DiVecchia:2021ndb,DiVecchia:2021bdo,Heissenberg:2021tzo}.

The eikonal phase at $\mathcal{O}(G^{n})$ is related to the $(D-2)$-dimensional Fourier transform of the $2\rightarrow2$ amplitude at the same order.
This latter quantity is
\begin{align}\label{eq:FT}
    \widetilde{\cM}_{n}(\mathbf{b})&=\frac{1}{4m_{1}m_{2}\sqrt{\omega^{2}-1}}\int\frac{d^{D-2}\mathbf{q}}{(2\pi)^{D-2}}e^{i\mathbf{q}\cdot\mathbf{b}}\cM_{n}(\mathbf{q}),
\end{align}
where $\omega\equiv v_{1}\cdot v_{2}$ is the product of the four-velocities of each particle, which have masses $m_{1,2}$.
We've also introduced the impact parameter $\mathbf{b}$ in the direction orthogonal to the asymptotic center-of-mass three-momentum.
Restricting the integrand of \cref{eq:FT} to the classical portion of the amplitude, this equation then defines the $n$PM eikonal phase,
\begin{align}
    \frac{1}{\hbar}\delta_{n}(\mathbf{b})=\widetilde{\cM}_{n}^{\text{cl.}}(\mathbf{b}),
\end{align}
with the entire eikonal phase being the sum of all $n$PM phases, $\delta(\mathbf{b})\equiv\sum_{n}\delta_{n}(\mathbf{b})$.

The exponentiation of the eikonal phase, in concert with a quantum remainder $\Delta$, describes the all-order, impact-parameter space, spinless amplitude \cite{DiVecchia:2021bdo,Heissenberg:2021tzo}:
\begin{align}\label{eq:AllOrderAmplitude}
    1+i\widetilde{\cM}(\mathbf{b})=\left[1+i\Delta(\mathbf{b})\right]e^{\frac{i}{\hbar}\delta(\mathbf{b})}.
\end{align}
We have absorbed factors of $2$ into $\Delta$ and $\delta$ relative to refs.~\cite{DiVecchia:2021bdo,Heissenberg:2021tzo}.
As its name suggests, the quantum remainder encapsulates all portions of the amplitude with a quantum $\hbar$ scaling.
The exponential of the eikonal phase, when expanded in powers of $G$, produces the part of the amplitude with a classical $\hbar$ scaling.
Moreover, products of $n$PM eikonal phases produce the so-called "super-classical" portions of the amplitude, which are singular as $\hbar\rightarrow0$.

Of particular relevance for our analysis here, expanding \cref{eq:AllOrderAmplitude}, the leading eikonal phase $\delta_{1}(\mathbf{b})$ and its square are given by the tree-level and super-classical one-loop amplitude as
\begin{align}
    \frac{i}{\hbar}\delta_{1}(\mathbf{b})&\equiv i\widetilde{\cM}_{1}^{\text{cl.}}(\mathbf{b})\\
    \frac{1}{2\hbar^{2}}\left[i\delta_{1}(\mathbf{b})\right]^{2}&=i\widetilde{\cM}_{2}^{\text{sc.}}(\mathbf{b}).\label{eq:LeadingEikonalSquared}
\end{align}
While the first of these is the definition of the leading eikonal, the second is dictated by the exponentiation of the eikonal.
\Cref{eq:LeadingEikonalSquared} has been verified for spinless scattering in general relativity and $\mathcal{N}=8$ supergravity \cite{KoemansCollado:2019ggb,Heissenberg:2021tzo}.
Our purpose here is to check whether it holds for the gravitational scattering of spinning particles.
In fact, we should expect \cref{eq:LeadingEikonalSquared} to be modified when spin is involved.
This is necessary so as to not produce spin structures outside of the solution space of the one-loop amplitude.
%These corrections are necessary because the left-hand side will contain terms existing outside of the Hilbert space of the right-hand side whenever spin is involved.
This point will be elucidated below.

We set $\hbar=1$ in the remainder of this note, but we classify the classicality of terms in impact-parameter space by counting powers of angular momentum.
By our definition of the impact parameter, it is related to the orbital angular momentum through $|\mathbf{b}|=|\mathbf{J}|/|\mathbf{p}|$, where $\mathbf{p}$ is the asymptotic center-of-mass three-momentum.
Also, a spin vector will scale with one power of the orbital angular momentum \cite{Bern:2020buy}.
Thus, classical terms at $\mathcal{O}(G)$ will scale as $|\mathbf{J}|^{4-D}$, while at $\mathcal{O}(G^{2})$ the classical scaling is $|\mathbf{J}|^{8-2D}$.
Terms with a quantum scaling have fewer powers of the angular momentum.

%\begin{align}
%    i\cM&=2s\int d^{D-2}\mathbf{b}\,e^{-i\mathbf{q}\cdot\mathbf{b}}\left(e^{i\delta(\mathbf{b})}-1\right).
%\end{align}

\section{The leading eikonal with spin}\label{sec:LeadingEikonal}

In this section we investigate, by direct computation, the exponentiation properties of the leading eikonal for spin-1/2 $\times$ spin-1/2 scattering up to $\mathcal{O}(G^{2})$.
We begin by deriving the leading eikonal with spin from the tree-level $2\rightarrow2$ amplitude.
Then, we provide a prescription for squaring the leading eikonal such that the square does not contain spin structures not in the solution space of spin-1/2 $\times$ spin-1/2 scattering.
Finally, we relate the square of the leading eikonal to the $\mathcal{O}(G^{2})$ super-classicalities in impact-parameter space.

\subsection{Deriving the leading eikonal}

\begin{figure}
    \centering
    \includegraphics{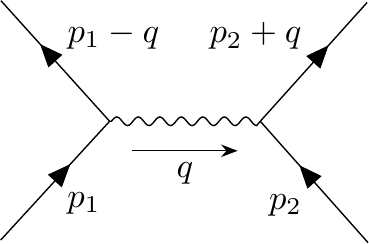}
    \caption{\label{fig:TreeLevel} The $2\rightarrow2$ diagram encoding classical effects at tree level.}
\end{figure}

As discussed above, the leading eikonal is simply related to the tree-level amplitude for $2\rightarrow2$ scattering, depicted in \cref{fig:TreeLevel}.
We work with heavy spin-$1/2$ states with masses $m_{i}$ and four-velocities $v_{i}^{\mu}$, which carry momenta $p_{i}^{\mu}=m_{i}v_{i}^{\mu}+k_{i}^{\mu}$ where $k_{i}^{\mu}$ are residual momenta scaling with $\hbar$ in the classical limit \cite{Damgaard:2019lfh}.
To compute the tree-level amplitude we need the Feynman rules for the three-point vertex and the graviton propagator:
\begin{subequations}
\begin{align}
    \includegraphics[valign=c]{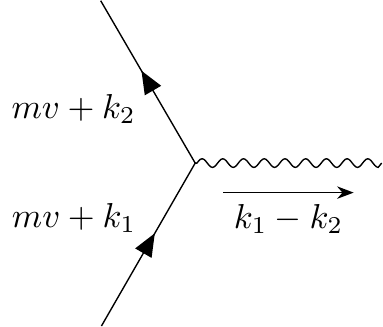}\,\,\mu\nu\quad&=\frac{i\kappa}{2}\left[-mv^{\mu}v^{\nu}+\frac{i}{4}(v^{\mu}\sigma^{\rho\nu}+v^{\nu}\sigma^{\rho\mu})(k_{2\rho}-k_{1\rho})\right]+\mathcal{O}(\hbar),\label{eq:ThreePointFRule} \\
    \mu\nu\,\,\includegraphics[valign=t]{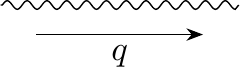}\,\,\alpha\beta\quad&=\frac{iP_{\mu\nu,\alpha\beta}}{q^{2}},
\end{align}
where $\kappa=\sqrt{32\pi G}$ and
\begin{align}
    P_{\mu\nu,\alpha\beta}\equiv\frac{1}{2}\left(\eta^{\mu\alpha}\eta^{\nu\beta}+\eta^{\mu\beta}\eta^{\nu\alpha}-\frac{2}{D-2}\eta^{\mu\nu}\eta^{\alpha\beta}\right),
\end{align}
for spacetime dimension $D=4-2\varepsilon$.
\end{subequations}

With these in hand, and setting $p_{i}^{\mu}=m_{i}v_{i}^{\mu}$, the tree-level amplitude is
\begin{align}\label{eq:CovariantTreeAmplitude}
    \cM_{1}^{1/2\times1/2}=-\frac{16\pi m_{1}^{2}m_{2}^{2}G}{q^{2}}&\left[\left(2\omega^{2}-\frac{2}{D-2}\right)\mathcal{U}_{1}\mathcal{U}_{2}+\frac{2i\omega}{m_{1}^{2}m_{2}}\mathcal{E}_{1}\mathcal{U}_{2}+\frac{2i\omega}{m_{1}m_{2}^{2}}\mathcal{U}_{1}\mathcal{E}_{2}\right.\notag \\
    %spinor identities in four dimensions
    &\qquad\qquad\left.+\frac{2\omega^{2}-1}{m_{1}m_{2}}(q\cdot S_{1}q\cdot S_{2}-q^{2}S_{1}\cdot S_{2})+\frac{2\omega}{m_{1}^{2}m_{2}^{2}}q^{2}p_{2}\cdot S_{1}p_{1}\cdot S_{2}\right],
    %some spinor identities in D dimensions
    %&\qquad\qquad\left.+\frac{D-2}{2}\frac{2\omega^{2}-\frac{2}{D-2}}{m_{1}m_{2}}(q\cdot S_{1}q\cdot S_{2}-q^{2}S_{1}\cdot S_{2})+\frac{(D-2)\omega}{m_{1}^{2}m_{2}^{2}}q^{2}p_{2}\cdot S_{1}p_{1}\cdot S_{2}\right],
\end{align}
where we have kept for now ultralocal terms with a classical $\hbar$ scaling.
The amplitude is normalized such that the spinors are dimensionless and satisfy the normalization condition $\bar{u}_{v}(p)u_{v}(p)=1$.
The subscript $v$ on the spinors denotes that we are using heavy spinors with velocity $v^{\mu}$.
The spin vector is defined in terms of the heavy spinors through
\begin{align}\label{eq:SpinDef}
    S^{\mu}_{i}&=\frac{1}{2}\bar{u}_{v_{i}}\gamma_{5}\gamma^{\mu}u_{v_{i}}.
\end{align}
We treat the spin and all spinor/gamma-matrix identities in four dimensions.
Finally, we've employed the shorthand notation
\begin{align}
    \mathcal{U}_{i}=\bar{u}_{v_{i}}u_{v_{i}},\quad \mathcal{E}_{i}=\epsilon_{\mu\nu\alpha\beta}p_{1}^{\mu}p_{2}^{\nu}q^{\alpha}S_{i}^{\beta}.
\end{align}

The leading eikonal is related to this amplitude in the center-of-mass frame through the two-dimensional Fourier transform \cref{eq:FT}.
Taking the initial momenta to be incoming, the center-of-mass kinemtics amount to \cite{Lorce:2017isp,Bern:2020buy}
\begin{align}\label{eq:CoMKinematics}
    p_{1}^{\mu}=(E_{1},\mathbf{p}),\quad p_{2}^{\mu}=(E_{2},-\mathbf{p}),&\quad q^{\mu}=(0,\mathbf{q}),\quad \mathbf{p}\cdot\mathbf{q}=\frac{\mathbf{q}^{2}}{2},\notag \\
    S_{1}^{\mu}=\left(\frac{\mathbf{p}\cdot\mathbf{S}_{1}}{m_{1}},\mathbf{S}_{1}+\frac{\mathbf{p}\cdot\mathbf{S}_{1}}{m_{1}(E_{1}+m_{1})}\mathbf{p}\right),&\quad S_{2}^{\mu}=\left(-\frac{\mathbf{p}\cdot\mathbf{S}_{2}}{m_{2}},\mathbf{S}_{2}+\frac{\mathbf{p}\cdot\mathbf{S}_{2}}{m_{2}(E_{2}+m_{2})}\mathbf{p}\right).
\end{align}
We have introduced the rest-frame spin vectors for each particle $\mathbf{S}_{i}$.
Substituting this into the amplitude, using $\mathcal{U}_{i}=1+\mathcal{O}(\hbar)$, and dropping all ultralocal terms (these terms yield Dirac deltas in impact-parameter space, and as such do not describe long-range interactions),
\begin{align}
    \cM_{1}^{1/2\times1/2}&=\frac{16\pi m_{1}^{2}m_{2}^{2}G}{\mathbf{q}^{2}}\left[\left(2\omega^{2}-\frac{2}{D-2}\right)+\frac{2E\omega}{m_{1}^{2}m_{2}}i(\mathbf{p}\times\mathbf{q})\cdot\mathbf{S}_{1}+\frac{2E\omega}{m_{1}^{2}m_{2}}i(\mathbf{p}\times\mathbf{q})\cdot\mathbf{S}_{2}+\frac{2\omega^{2}-1}{m_{1}m_{2}}\mathbf{q}\cdot\mathbf{S}_{1}\mathbf{q}\cdot\mathbf{S}_{2}\right],
\end{align}
where $E=E_{1}+E_{2}$ is the total energy.
Finally, having the amplitude in the center-of-mass frame, we can find the leading eikonal:
\begin{align}\label{eq:LeadingEikonal}
    i\delta_{1}&=i\widetilde{\cM}_{1}^{1/2\times1/2}=\frac{i\pi^{2-D/2}Gm_{1}m_{2}}{\sqrt{\omega^{2}-1}}\left[\left(2\omega^{2}-\frac{2}{D-2}\right)\frac{\Gamma(D/2-2)}{\mathbf{b}^{D-4}}-\frac{4E\omega}{m_{1}m_{2}}\frac{\Gamma(D/2-1)}{\mathbf{b}^{D-2}}(\mathbf{p}\times\mathbf{b})\cdot\mathbf{a}_{1}\right.\notag \\
    &\quad\left.-\frac{4E\omega}{m_{1}m_{2}}\frac{\Gamma(D/2-1)}{\mathbf{b}^{D-2}}(\mathbf{p}\times\mathbf{b})\cdot\mathbf{a}_{2}+2(2\omega^{2}-1)\frac{\Gamma(D/2-1)}{\mathbf{b}^{D-2}}\left(\Pi^{ij}-(D-2)\frac{\mathbf{b}^{i}\mathbf{b}^{j}}{\mathbf{b}^{2}}\right)\mathbf{a}_{1}^{i}\mathbf{a}_{2}^{j}\right],
\end{align}
where $\mathbf{a}_{i}\equiv\mathbf{S}_{i}/m_{i}$.
The projector $\Pi^{ij}$ onto the plane orthogonal to $\mathbf{p}$ is defined in \cref{eq:Projector}.
Note that in order to obtain the spin-monopole term of the leading eikonal (which is in agreement with ref.~\cite{KoemansCollado:2019ggb}) we had to use $\mathcal{U}_{i}=1+\mathcal{O}(\hbar)$.
Implicit in this identity is that the polarizations of the spin-1/2 particles are unchanged in the scattering.

For convenience later on, let us introduce the notation $X^{A}$, labelling a specific spin-structure portion of the quantity $X$.
The different values of $A$ and the corresponding spin structures are
\begin{align*}
    A=(0)\rightarrow 1,\quad A&=(1,1)\rightarrow(\mathbf{p}\times\mathbf{q})\cdot\mathbf{S}_{1},\quad A=(1,2)\rightarrow(\mathbf{p}\times\mathbf{q})\cdot\mathbf{S}_{2}, \\
    A=(2,1)\rightarrow\mathbf{q}\cdot\mathbf{S}_{1}\mathbf{q}\cdot\mathbf{S}_{2},\quad A&=(2,2)\rightarrow \mathbf{q}^{2}\mathbf{S}_{1}\cdot\mathbf{S}_{2},\quad A=(2,3)\rightarrow\mathbf{q}^{2}\mathbf{p}\cdot\mathbf{S}_{1}\mathbf{p}\cdot\mathbf{S}_{2}.
\end{align*}
This notation was first employed in ref.~\cite{Bern:2020buy}.

Before moving on, let us remark that the leading eikonal in \cref{eq:LeadingEikonal} produces the known aligned-spin scattering angle when $D\rightarrow4$ \cite{Vines:2017hyw}.

\subsection{Squaring the spinning eikonal}\label{sec:SquaringEikonal}

Directly squaring \cref{eq:LeadingEikonal} can be seen to produce spin structures that are outside the solution space of a spin-1/2 $\times$ spin-1/2 amplitude.
Specifically, such an amplitude can only contain effects up to linear order in the spin of each particle, while squaring \cref{eq:LeadingEikonal} will yield terms of the schematic form $S_{1}^{2},\,S_{2}^{2},\,S_{1}S_{2}^{2},\,S_{1}^{2}S_{2},$ and $S_{1}^{2}S_{2}^{2}$.
If the square of the leading eikonal is to be comparable to a one-loop, spin-1/2 $\times$ spin-1/2 amplitude, these terms must be removed from the square.

One way to ensure that the square does not contain these structures is to redefine how the eikonal is squared.
At leading order in $\hbar$, the external polarizations are independent of the transfer momentum,\footnote{We have used this implicitly in \Cref{sec:LeadingEikonal} when we wrote $\mathcal{U}_{i}=1+\mathcal{O}(\hbar)$. This is simply a consequence of boosting the final-state spinors to have the incoming momenta.} and are therefore inert under the Fourier transform \cref{eq:FT}.
We can thus evaluate the leading eikonal with external polarizations present.
With polarizations, we define the square of the eikonal (with polarization labels $s$) as
\begin{align}\label{eq:SquaringEikonal}
    \left[\delta_{1}(\mathbf{b};s\rightarrow s)\right]^{2}\rightarrow\delta_{1}(\mathbf{b};s\rightarrow s^{\prime})\otimes\delta_{1}(\mathbf{b};s^{\prime}\rightarrow s)\equiv\sum_{s^{\prime}}\delta_{1}(\mathbf{b};s\rightarrow s^{\prime})\delta_{1}(\mathbf{b};s^{\prime}\rightarrow s).
\end{align}
Again, all external states involved in this product depend only on the incoming momenta.
There must be a polarization sum for each spinning particle involved in the scattering, though for brevity we have only explicitly shown one sum.

Specializing to the spin-1/2 case, as we've expressed the eikonal in terms of the rest frame spin, we will restore the rest frame heavy spinors to \cref{eq:LeadingEikonal} instead of the relativistic heavy spinors.
For spin-monopole factors, this amounts to restoring $\bar{\xi}^{s}_{v}\xi^{s^{\prime}}_{v}=\delta^{ss^{\prime}}$.
For spin contributions, the expressions in \cref{eq:CoMKinematics} for the covariant spin demonstrate that the rest frame spin vector is simply the spatial component of the covariant spin when the particle is at rest.
Then, in the Weyl representation, \cref{eq:SpinDef} becomes
\begin{align}
    \mathbf{S}^{ss^{\prime}}=\frac{1}{2}\bar{\xi}_{v}^{s}\gamma_{5}\gamma^{i}\xi_{v}^{s^{\prime}}=\frac{1}{2}\bar{\xi}^{s}_{v}\begin{pmatrix}
    0 & \vec{\sigma} \\
    \vec{\sigma} & 0
    \end{pmatrix}\xi^{s^{\prime}}_{v}.
\end{align}
Here, $\vec{\sigma}$ is the Pauli-matrix three-vector.

The polarization sum in \cref{eq:SquaringEikonal} now makes spin structures outside the solution space of spin-1/2 $\times$ spin-1/2 scattering subleading in $\hbar$.
Let us work this out explicitly for spin-1/2 particles.
Heavy spinors are related to standard Dirac spinors (normalized such that $\bar{u}(p)u(p)=1$) simply through \cite{Georgi:1990um,Damgaard:2019lfh}
\begin{align}\label{eq:HeavySpinorDef}
    u_{v}(p;s)=\frac{1+\slashed{v}}{2}u(p;s),
\end{align}
for momentum $p^{\mu}=mv^{\mu}+l^{\mu}$, where $v^{2}=1$ and $l\sim\hbar$.
This allows us to easily evaluate the polarization sum for heavy spinors:
\begin{align}
    \sum_{s}u_{v}(p;s)\bar{u}_{v}(p;s)=\frac{1+\slashed{v}}{2}\left[\sum_{s}u(p;s)\bar{u}(p;s)\right]\frac{1+\slashed{v}}{2}=\frac{1+\slashed{v}}{2}\frac{\slashed{p}+m}{2m}\frac{1+\slashed{v}}{2}=\frac{1+\slashed{v}}{2}+\mathcal{O}(\hbar^{2}).
\end{align}
We have used here the on-shell condition $v\cdot l=-l^{2}/2m$.
Notably, the polarization sum up to this order is entirely independent of the residual momentum $l^{\mu}$.
By definition, the rest frame spinors are the $p\rightarrow m(1,\vec{0})$ limit of the relativistic spinors.
We thus find the completeness relation
\begin{align}
    \sum_{s}\xi_{v}^{s}\bar{\xi}_{v}^{s}=\frac{1+\gamma^{0}}{2}.
\end{align}
Considering now a term in the square that is quadratic in the spin of, say, particle 1,
\begin{align}
    \sum_{s^{\prime}}\mathbf{S}^{i,ss^{\prime}}_{1}\mathbf{S}^{j,s^{\prime}s}_{1}=\frac{1}{4}\bar{\xi}_{v_{1}}^{s}\begin{pmatrix}
    0 & \sigma^{i} \\
    \sigma^{i} & 0
    \end{pmatrix}\frac{1+\gamma^{0}}{2}\begin{pmatrix}
    0 & \sigma^{j} \\
    \sigma^{j} & 0
    \end{pmatrix}\xi_{v_{1}}^{s}=\frac{1}{8}\bar{\xi}_{v_{1}}^{s}\begin{pmatrix}
    \sigma^{i}\sigma^{j} & \sigma^{i}\sigma^{j} \\
    \sigma^{i}\sigma^{j} & \sigma^{i}\sigma^{j}
    \end{pmatrix}\xi_{v_{1}}^{s}.
\end{align}
The product of sigma matrices is $\sigma^{i}\sigma^{j}=\delta^{ij}+i\epsilon^{ijk}\sigma^{l}$.
Since we have converted two powers of the spin in a term with classical scaling to terms with one or no powers of the spin, the result of this reduction is contributions that have a quantum scaling, and can therefore be ignored.
The fact that $\frac{1}{2}(1+\gamma^{0})\xi_{v}^{s}=\xi_{v}^{s}$\footnote{This is simply the Dirac equation in the rest frame. Alternatively, it can be seen from the definition of the heavy spinor in \cref{eq:HeavySpinorDef}.} means that all other structures that are linear in the spin of each spin-1/2 particle are unaffected by the polarization sum. 

We have shown here the squaring procedure for the spin-1/2 $\times$ spin-1/2 case.
For particles with other spins one must employ the polarization sum accordingly.
In the case of scalar-scalar scattering, the polarization sum is trivial and \cref{eq:SquaringEikonal} reduces to a regular product.

\subsection{The leading eikonal at one loop}

\begin{figure}
\centering
\subfloat{
    \includegraphics[scale=0.4]{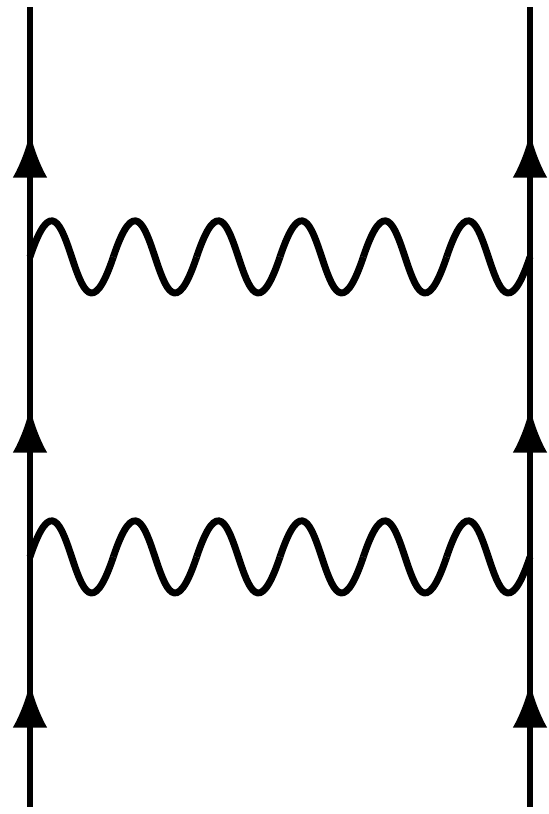}
    }
\hspace{2cm}
\subfloat{
    \includegraphics[scale=0.4]{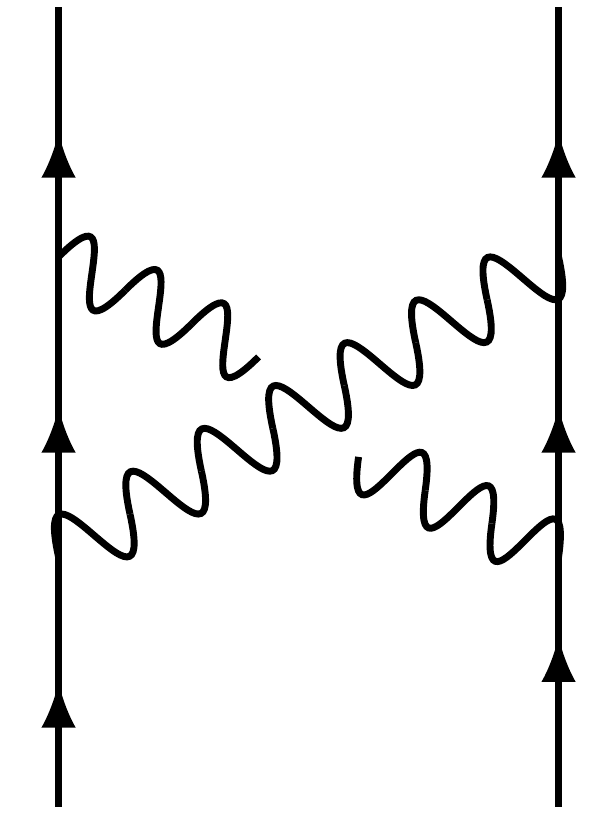}
    }
\caption{\label{fig:Boxes}The box and cross-box topologies. These diagrams yield the dominant contributions to the one-loop amplitude in the classical limit.}
\end{figure}

In the spinless case it is well understood that the exponentiation of the leading eikonal generates the dominant portions of higher-loop amplitudes in the classical limit \cite{Levy:1969cr,DiVecchia:2019myk,DiVecchia:2021bdo,Heissenberg:2021tzo}.
Practically, at one-loop order, this means that \cref{eq:LeadingEikonalSquared} relates the square of the leading eikonal to the leading contributions from the box and cross-box diagrams; see \cref{fig:Boxes}.
Let us establish whether a similar relation holds when spin is included.
In order to investigate this, we now compute the leading box and cross-box contributions to the $2\rightarrow2$ amplitude for the scattering of two spin-$1/2$ particles at $\mathcal{O}(G^{2})$.
Upon tensor reducing higher-rank loop integrals, we find that the super-classical portion of the amplitude can be written as
\begin{align}\label{eq:OneLoopSchem}
    i\mathcal{M}_{2}^{\text{sc.}}&=iM_{2}^{\text{sc.}}\left(\mathcal{I}_{\square}+\mathcal{I}_{\boxtimes}\right),
\end{align}
where $\mathcal{I}_{\square,\boxtimes}$ are the scalar integrals corresponding to the topologies in \cref{fig:Boxes}.
The sum of their values is given in \Cref{sec:LoopIntegrals}.
Here, $iM^{\text{sc.}}$ is the remaining integrand after tensor reduction of the integrals, and as such is independent of the loop momentum.
Also, we have dropped the superscript labelling the spins of the scattering particles, understanding that two spin-$1/2$ particles are scattered in this section.
For clarity, let us consider \cref{eq:OneLoopSchem} at each order in spin independently.
In the proceeding subsections, we use the square of the leading eikonal to mean \cref{eq:SquaringEikonal}.
All amplitudes computed below agree with the super-classical pieces computed in ref.~\cite{Damgaard:2019lfh} in the limit $D\rightarrow4$.

\subsubsection{Spin monopole}

Expressed directly in the center-of-mass frame, the spin-monopole portion is
\begin{align}
    i\cM^{\text{sc.},(0)}_{2}    &=-\frac{\pi^{3-D/2}G^{2}m_{1}^{3}m_{2}^{3}}{2^{D-7}\sqrt{\omega^{2}-1}}\left(2\omega^{2}-\frac{2}{D-2}\right)^{2}\frac{\Gamma(D/2-2)^{2}\Gamma(3-D/2)}{\Gamma(D-4)}\left(\mathbf{q}^{2}\right)^{D/2-3}.
\end{align}
This portion is already known to exponentiate in impact-parameter space; see e.g. ref.~\cite{KoemansCollado:2019ggb}.
Nevertheless, for completeness, we present the Fourier transform of this portion:
\begin{align}
    i\widetilde{\cM}_{2}^{\text{sc.},(0)}=-\frac{\pi^{4-D}G^{2}m_{1}^{2}m_{2}^{2}}{2(\omega^{2}-1)}\left(2\omega^{2}-\frac{2}{D-2}\right)^{2}\frac{\Gamma(D/2-2)^{2}}{\mathbf{b}^{2D-8}},
\end{align}
in agreement with the spin-monopole portion of $\frac{1}{2}(i\delta_{1})^{2}$.

\subsubsection{Linear in spin}

Moving on, the spin-dipole portion of the amplitude in the center-of-mass frame is
\begin{align}
    i\cM_{2}^{\text{sc.},(1,1)}&=-\frac{\pi^{3-D/2}G^{2}m_{1}m_{2}^{2}}{2^{D-8}\sqrt{\omega^{2}-1}}E\omega\left(2\omega^{2}-\frac{2}{D-2}\right)\frac{\Gamma(D/2-2)^{2}\Gamma(3-D/2)}{\Gamma(D-4)}i(\mathbf{p}\times\mathbf{q})\cdot\mathbf{S}_{1}\left(\mathbf{q}^{2}\right)^{D/2-3}.
\end{align}
In impact-parameter space,
\begin{align}
    i\widetilde{\cM}_{2}^{\text{sc.},(1,1)}&=\frac{4\pi^{4-D}G^{2}m_{1}m_{2}}{(\omega^{2}-1)}\frac{\Gamma(D/2-2)\Gamma(D/2-1)}{\mathbf{b}^{2D-6}}E\omega\left(2\omega^{2}-\frac{2}{D-2}\right)(\mathbf{p}\times\mathbf{b})\cdot\mathbf{a}_{1}.
\end{align}
This agrees precisely with the spin-dipole portion of $\frac{1}{2}(i\delta_{1})^{2}$, namely $\frac{1}{2}(i\delta_{1}^{(0)})\otimes(i\delta_{1}^{(1,1)})+\frac{1}{2}(i\delta_{1}^{(1,1)})\otimes(i\delta_{1}^{(0)})$.
%This provides the first -- to the best of our knowledge -- evidence for the eikonal exponentiation of linear-in-spin effects.
The analogous result holds for the spin of the other particle by swapping the labels $1\leftrightarrow2$.

\subsubsection{Quadratic in spin}

Finally, we consider the quadratic-in-spin portion of the amplitude.
It will be clearer to divide our analysis further, studying each spin structure independently.

Consider first the $\mathbf{q}\cdot\mathbf{S}_{1}\mathbf{q}\cdot\mathbf{S}_{2}$ portion of the amplitude.
This is
%spin identities in four dimensions
\begin{align}
    i\cM^{\text{sc.},(2,1)}_{2}&=-\frac{\pi^{3-D/2}G^{2}m_{1}^{2}m_{2}^{2}}{2^{D-7}\sqrt{\omega^{2}-1}}\frac{\Gamma(D/2-2)^{2}\Gamma(3-D/2)}{\Gamma(D-4)}\mathbf{q}\cdot\mathbf{S}_{1}\mathbf{q}\cdot\mathbf{S}_{2}\left(\mathbf{q}^{2}\right)^{D/2-3}\frac{4(D-3)\omega^{4}+(8-3D)\omega^{2}+1}{(D-3)}.
\end{align}
%some spin identities in D dimensions
%\begin{align}
%    i\cM^{\text{sc.},(2,1)}_{2}&=-\frac{\pi^{3-D/2}G^{2}m_{1}^{2}m_{2}^{2}}{2^{D-7}\sqrt{\omega^{2}-1}}\frac{\Gamma(D/2-2)^{2}\Gamma(3-D/2)}{\Gamma(D-4)}\mathbf{q}\cdot\mathbf{S}_{1}\mathbf{q}\cdot\mathbf{S}_{2}\left(\mathbf{q}^{2}\right)^{D/2-3}\notag \\
%    &\quad\times\frac{(D^{2}-2D-4)\omega^{4}-4(D-3)\omega^{2}+1}{(D-3)}.
%\end{align}
As before, we convert this to impact-parameter space in order to compare with the square of the leading eikonal.
The Fourier transform in this case will contain all quadratic-in-spin structures, not just the $(2,1)$ structure.
Keeping all resulting structures, the Fourier transform is
%spin identities in four dimensions
\begin{align}\label{eq:21LoopAmp}
    &i\widetilde{\cM}_{2}^{\text{sc.},(2,1)}+i\widetilde{\cM}_{2}^{\text{sc.},(2,1\rightarrow2)}+i\widetilde{\cM}_{2}^{\text{sc.},(2,1\rightarrow3)}\notag \\
    &=-\frac{2\pi^{4-D}G^{2}m_{1}^{2}m_{2}^{2}}{(\omega^{2}-1)}\frac{\Gamma(D/2-2)\Gamma(D/2-1)}{\mathbf{b}^{2D-6}}\left(-2\frac{\mathbf{b}\cdot\mathbf{a}_{1}\mathbf{b}\cdot\mathbf{a}_{2}}{\mathbf{b}^{2}}+\frac{\Pi^{ij}\mathbf{a}^{i}_{1}\mathbf{a}^{j}_{2}}{D-3}\right)\left[4(D-3)\omega^{4}+(8-3D)\omega^{2}+1\right].
\end{align}
%some spin identitites in D dimensions
%\begin{align}\label{eq:21LoopAmp}
%    &i\widetilde{\cM}_{2}^{\text{sc.},(2,1)}+i\widetilde{\cM}_{2}^{\text{sc.},(2,1\rightarrow2)}+i\widetilde{\cM}_{2}^{\text{sc.},(2,1\rightarrow3)}\notag \\
%    &=-\frac{2\pi^{4-D}G^{2}m_{1}^{2}m_{2}^{2}}{(\omega^{2}-1)}\frac{\Gamma(D/2-2)\Gamma(D/2-1)}{\mathbf{b}^{2D-6}}\left(2\frac{\mathbf{b}\cdot\mathbf{a}_{1}\mathbf{b}\cdot\mathbf{a}_{2}}{\mathbf{b}^{2}}+\frac{\Pi^{ij}\mathbf{a}^{i}_{1}\mathbf{a}^{j}_{2}}{D-3}\right)\left[(D^{2}-2D-4)\omega^{4}-4(D-3)\omega^{2}+1\right].
%\end{align}
The second term in the large round brackets contains the spill-over into the $(2,2)$ and $(2,3)$ portions of the impact-parameter space amplitude.

Focusing on the first term in the large round brackets for now, this is to be compared with the $(2,1)$ spin structure from the square of the leading eikonal:
\begin{align}\label{eq:21EikSquared}
    \left[\frac{1}{2}\left(i\delta_{1}\right)^{2}\right]^{(2,1)}&=\frac{2\pi^{4-D}G^{2}m_{1}^{2}m_{2}^{2}}{(\omega^{2}-1)}\frac{\Gamma(D/2-2)\Gamma(D/2-1)}{\mathbf{b}^{2D-4}}\mathbf{b}\cdot\mathbf{a}_{1}\mathbf{b}\cdot\mathbf{a}_{2}\notag \\
    &\quad\times\left[(D-2)\left(2\omega^{2}-1\right)\left(2\omega^{2}-\frac{2}{D-2}\right)+(D-4)4\omega^{2}(\omega^{2}-1)\right].
\end{align}
Simplifying, we find that the $(2,1)$ portion of \cref{eq:21LoopAmp} is in agreement with \cref{eq:21EikSquared}:
\begin{align}
    i\widetilde{\cM}_{2}^{(2,1)}&=\left[\frac{1}{2}\left(i\delta_{1}\right)^{2}\right]^{(2,1)}.
\end{align}

Next, we investigate the $\mathbf{q}^{2}\mathbf{S}_{1}\cdot\mathbf{S}_{2}$ terms.
We find the leading part of the amplitude to be
%spin identities in four dimensions
\begin{align}
    i\cM^{\text{sc.},(2,2)}_{2}&=\frac{\pi^{3-D/2}G^{2}m_{1}^{2}m_{2}^{2}}{2^{D-7}\sqrt{\omega^{2}-1}}\frac{\Gamma(D/2-2)^{2}\Gamma(3-D/2)}{\Gamma(D-4)}\mathbf{S}_{1}\cdot\mathbf{S}_{2}\left(\mathbf{q}^{2}\right)^{D/2-2}\frac{4(D-2)\omega^{4}+(4-3D)\omega^{2}+1}{(D-2)}.
\end{align}
%some spin identities in d dimensions
%\begin{align}
%    i\cM^{\text{sc.},(2,2)}_{2}&=\frac{\pi^{3-D/2}G^{2}m_{1}^{2}m_{2}^{2}}{2^{D-7}\sqrt{\omega^{2}-1}}\frac{\Gamma(D/2-2)^{2}\Gamma(3-D/2)}{\Gamma(D-4)}\mathbf{S}_{1}\cdot\mathbf{S}_{2}\left(\mathbf{q}^{2}\right)^{D/2-2}\notag \\
%    &\quad\times\frac{(D^{2}-12)(D-2)\omega^{4}+(-5D^{2}+25D-28)\omega^{2}+1}{(D-2)(D-3)}.
%\end{align}
In impact-parameter space we must combine the Fourier transform of this with the contribution $i\widetilde{\cM}^{(2,1\rightarrow2)}_{2}$ from the transform of the $(2,1)$ spin structure:
%spin identities in four dimensions
\begin{align}\label{eq:22LoopAmp}
    &i\widetilde{\cM}^{\text{sc.},(2,2)}_{2}+i\widetilde{\cM}_{2}^{\text{sc.},(2,1\rightarrow2)}=-\frac{2\pi^{4-D}G^{2}m_{1}^{2}m_{2}^{2}}{(\omega^{2}-1)}\frac{\Gamma(D/2-2)\Gamma(D/2-1)}{\mathbf{b}^{2D-6}}\mathbf{a}_{1}\cdot\mathbf{a}_{2}\notag \\
    &\times\left[\frac{4(D-2)\omega^{4}+(4-3D)\omega^{2}+1}{(D-2)}(D-4)+\frac{4(D-3)\omega^{4}+(8-3D)\omega^{2}+1}{(D-3)}\right].
\end{align}
%some spin identities in d dimensions
%\begin{align}\label{eq:22LoopAmp}
%    &i\widetilde{\cM}^{\text{sc.},(2,2)}_{2}+i\widetilde{\cM}_{2}^{\text{sc.},(2,1\rightarrow2)}=-\frac{2\pi^{4-D}G^{2}m_{1}^{2}m_{2}^{2}}{(\omega^{2}-1)}\frac{\Gamma(D/2-2)\Gamma(D/2-1)}{\mathbf{b}^{2D-6}}\mathbf{a}_{1}\cdot\mathbf{a}_{2}\notag \\
%    &\times\left[\frac{(D^{2}-12)(D-2)\omega^{4}+(-5D^{2}+25D-28)\omega^{2}+1}{(D-2)(D-3)}(D-4)+\frac{(D^{2}-2D-4)\omega^{4}+4(D-3)\omega^{2}+1}{(D-3)}\right].
%\end{align}
Note that the direct contribution $i\widetilde{\cM}^{(2,2)}_{2}$ is finite in the limit $D\rightarrow 4$, in constrast to all other terms in impact-parameter space derived so far.
This is a consequence of the additional factor of $\mathbf{q}^{2}$ coming from the spin structure.

To scrutinize the potential exponential structure of the leading eikonal, \cref{eq:22LoopAmp} is to be compared against $[\frac{1}{2}(i\delta_{1})^{2}]^{(2,2)}$.
This structure is
\begin{align}\label{eq:22EikSquared}
    \left[\frac{1}{2}\left(i\delta_{1}\right)^{2}\right]^{(2,2)}&=-\frac{2\pi^{4-D}G^{2}m_{1}^{2}m_{2}^{2}}{(\omega^{2}-1)}\frac{\Gamma(D/2-2)\Gamma(D/2-1)}{\mathbf{b}^{2D-6}}\mathbf{a}_{1}\cdot\mathbf{a}_{2}\notag \\
    %spin identities in four dimensions
    &\quad\times\left[\left(2\omega^{2}-1\right)\left(2\omega^{2}-\frac{2}{D-2}\right)+4\omega^{2}(\omega^{2}-1)(D-4)\right].
    %some spin identities in d dimensions
    %&\quad\times\left[\frac{D-2}{2}\left(2\omega^{2}-\frac{2}{D-2}\right)^{2}+4\omega^{2}(\omega^{2}-1)(D-4)\right].
\end{align}
In general spacetime dimensions \cref{eq:22LoopAmp,eq:22EikSquared} are not equivalent.
However, they are related in the limit $D\rightarrow4$:
\begin{align}
    \lim_{D\rightarrow4}\left(i\widetilde{\cM}^{\text{sc.},(2,2)}_{2}+i\widetilde{\cM}_{2}^{\text{sc.},(2,1\rightarrow2)}\right)&=\lim_{D\rightarrow4}\left[\frac{1}{2}\left(i\delta_{1}\right)^{2}\right]^{(2,2)},
\end{align}
up to terms that vanish in this limit.
We will return to a discussion of this limit at the end of this section, and in our conclusions.

The final structure to investigate is the $\mathbf{q}^{2}\mathbf{p}\cdot\mathbf{S}_{1}\mathbf{p}\cdot\mathbf{S}_{2}$ portion of the amplitude.
In the center-of-mass frame this is
%spin identities in four dimensions
\begin{align}
    &i\cM^{\text{sc.},(2,3)}_{2}=\frac{\pi^{3-D/2}G^{2}m_{1}^{2}m_{2}^{2}}{2^{D-7}\sqrt{\omega^{2}-1}}\frac{\Gamma(D/2-2)^{2}\Gamma(3-D/2)}{\Gamma(D-4)}\left(\mathbf{q}^{2}\right)^{D/2-2}\frac{\mathbf{p}\cdot\mathbf{S}_{1}\mathbf{p}\cdot\mathbf{S}_{2}}{\mathbf{p}^{2}}\notag \\
    %&\quad\times\left[-\frac{\omega}{\mathbf{p}^{2}}a_{2}^{(2,3)}+M_{12}a_{2}^{(2,2)}\right].
    &\quad\times\left[-\omega\frac{4(D-2)(D-3)\omega^{4}+(-4D^{2}+19D-20)\omega^{2}+2D-7}{(D-3)(D-2)}+(\omega-1)\frac{4(D-2)\omega^{4}+(4-3D)\omega^{2}+1}{D-2}\right].
\end{align}
%some spin identities in d dimensions
%\begin{align}
%    &i\cM^{\text{sc.},(2,3)}_{2}=\frac{\pi^{3-D/2}G^{2}m_{1}^{2}m_{2}^{2}}{2^{D-7}\sqrt{\omega^{2}-1}}\frac{\Gamma(D/2-2)^{2}\Gamma(3-D/2)}{\Gamma(D-4)}\left(\mathbf{q}^{2}\right)^{D/2-2}\frac{\mathbf{p}\cdot\mathbf{S}_{1}\mathbf{p}\cdot\mathbf{S}_{2}}{\mathbf{p}^{2}}\notag \\
%    %&\quad\times\left[-\frac{\omega}{\mathbf{p}^{2}}a_{2}^{(2,3)}+M_{12}a_{2}^{(2,2)}\right].\\
%    &\quad\times\left[\frac{-(D^{2}-12)(D-2)\omega^{4}+(D-4)(D-2)(D-1)\omega^{3}+(5D^{2}-25D+28)\omega^{2}-(D-4)(D-1)\omega-1}{(D-3)(D-2)}\right].
%\end{align}
%\textbf{(Investigating the one-loop result in ref.~\cite{Bern:2020buy} it looks like $M_{12}=(\omega-1)/\mathbf{p}^{2}$. I confirmed this.)}
When converting this to impact-parameter space, we must again account for the correction from the Fourier transform of the $(2,1)$ spin structure.
All-in-all we find
%spin identities in four dimensions
\begin{align}\label{eq:23LoopAmp}
    &i\widetilde{\cM}^{\text{sc.},(2,3)}_{2}+i\widetilde{\cM}^{\text{sc.},(2,1\rightarrow3)}_{2}=-\frac{2\pi^{4-D}G^{2}m_{1}^{2}m_{2}^{2}}{(\omega^{2}-1)}\frac{\Gamma(D/2-2)\Gamma(D/2-1)}{\mathbf{b}^{2D-6}}\frac{\mathbf{p}\cdot\mathbf{a}_{1}\mathbf{p}\cdot\mathbf{a}_{2}}{\mathbf{p}^{2}}\notag \\
    &\times\left[(D-4)\left(-\omega\frac{4(D-2)(D-3)\omega^{4}+(-4D^{2}+19D-20)\omega^{2}+2D-7}{(D-3)(D-2)}+(\omega-1)\frac{4(D-2)\omega^{4}+(4-3D)\omega^{2}+1}{D-2}\right)\right.\notag \\
    &\qquad\qquad\left.-\frac{4(D-3)\omega^{4}+(8-3D)\omega^{2}+1}{(D-3)}\right].
\end{align}
%some spin identities in d dimensions
%\begin{align}\label{eq:23LoopAmp}
%    &i\widetilde{\cM}^{\text{sc.},(2,3)}_{2}+i\widetilde{\cM}^{\text{sc.},(2,1\rightarrow3)}_{2}=-\frac{2\pi^{4-D}G^{2}m_{1}^{2}m_{2}^{2}}{(\omega^{2}-1)}\frac{\Gamma(D/2-2)\Gamma(D/2-1)}{\mathbf{b}^{2D-6}}\frac{\mathbf{p}\cdot\mathbf{a}_{1}\mathbf{p}\cdot\mathbf{a}_{2}}{\mathbf{p}^{2}}\notag \\
%    &\times\left[(D-4)\frac{-(D^{2}-12)(D-2)\omega^{4}+(D-4)(D-2)(D-1)\omega^{3}+(5D^{2}-25D+28)\omega^{2}-(D-4)(D-1)\omega-1}{(D-3)(D-2)}\right.\notag \\
%    &\qquad\qquad\left.-\frac{(D^{2}-2D-4)\omega^{4}+4(D-3)\omega^{2}+1}{(D-3)}\right].
%\end{align}
The overall $\mathbf{q}^{2}$ associated with this spin structure again makes it so that the direct contribution $i\widetilde{\cM}^{\text{sc.},(2,3)}_{2}$ is finite for $D\rightarrow4$.
Comparing to the appropriate portion of the square of the leading eikonal,
\begin{align}
    \left[\frac{1}{2}\left(i\delta_{1}\right)^{2}\right]^{(2,3)}&=\frac{2\pi^{4-D}G^{2}m_{1}^{2}m_{2}^{2}}{(\omega^{2}-1)}\frac{\Gamma(D/2-2)\Gamma(D/2-1)}{\mathbf{b}^{2D-6}}\frac{\mathbf{p}\cdot\mathbf{a}_{1}\mathbf{p}\cdot\mathbf{a}_{2}}{\mathbf{p}^{2}}\notag \\
    &\quad\times\left[\left(2\omega^{2}-1\right)\left(2\omega^{2}-\frac{2}{D-2}\right)+(D-4)4\omega^{2}(\omega^{2}-1)\right],
\end{align}
once again we find a mismatch in general dimensions --- and once again a relation between the two quantities exists in the $D\rightarrow4$ limit:
\begin{align}
    \lim_{D\rightarrow4}\left(i\widetilde{\cM}^{\text{sc.},(2,3)}_{2}+i\widetilde{\cM}^{\text{sc.},(2,1\rightarrow3)}_{2}\right)&=\lim_{D\rightarrow4}\left[\frac{1}{2}\left(i\delta_{1}\right)^{2}\right]^{(2,3)}.
\end{align}
%\begin{align}
%    \left.\lim_{D\rightarrow4}\left(i\widetilde{\cM}^{(2,3)}_{2}+i\widetilde{\cM}^{(2,1\rightarrow3)}_{2}\right)\right|_{\text{div+n.a.}}&=\left.\lim_{D\rightarrow4}\left[\frac{1}{2}\left(i\delta_{1}\right)^{2}\right]^{(2,3)}\right|_{\text{div.+n.a.}}\notag \\
%    &=\frac{4G^{2}m_{1}^{2}m_{2}^{2}}{(\omega^{2}-1)}\frac{(2\omega^{2}-1)^{2}}{\mathbf{b}^{2}}\left(\frac{1}{D-4}-\log\mathbf{b}^{2}\right)\frac{\mathbf{p}\cdot\mathbf{S}_{1}\mathbf{p}\cdot\mathbf{S}_{2}}{\mathbf{p}^{2}}.
%\end{align}

Let us summarize what we have seen here, labelling the entire amplitude in impact-parameter space as $i\chi_{2}\equiv i\widetilde{\cM}_{2}^{\text{sc.}}$.
We found the following relations between the different spin structures of $i\chi_{2}$ and the square of the leading eikonal:
\begin{subequations}\label{eq:EikExp}
\begin{align}
    i\chi_{2}^{(0)}&=\left[\frac{1}{2}\left(i\delta_{1}\right)^{2}\right]^{(0)}, \\
    i\chi^{(1,i)}_{2}=\left[\frac{1}{2}\left(i\delta_{1}\right)^{2}\right]^{(1,i)},&\quad i\chi_{2}^{(2,1)}=\left[\frac{1}{2}\left(i\delta_{1}\right)^{2}\right]^{(2,1)}, \\
    \lim_{D\rightarrow4}i\chi_{2}^{(2,2)}&=\lim_{D\rightarrow4}\left[\frac{1}{2}\left(i\delta_{1}\right)^{2}\right]^{(2,2)}, \\ \lim_{D\rightarrow4}i\chi_{2}^{(2,3)}&=\lim_{D\rightarrow4}\left[\frac{1}{2}\left(i\delta_{1}\right)^{2}\right]^{(2,3)}.
\end{align}
\end{subequations}
We have thus found evidence for the exponentiation of the leading eikonal, with the caveat that the exponentiation takes place only in the limit where $D\rightarrow4$.
The equalities in the last two lines of \cref{eq:EikExp} hold for the divergent and finite parts in the limit $D\rightarrow4$, and disagree on terms that vanish in this limit.
%A potential explanation for the mismatch observed in general dimensions for the $(2,2)$ and $(2,3)$ spin structures is that, at $\mathcal{O}(G)$, they only arise after converting the $(2,1)$ structure to impact-parameter space.
%This is in contrast to the amplitude at $\mathcal{O}(G^{2})$, which already contains local terms proportional to these two spin structures.
%Nevertheless, it is remarkable that this is not an obstacle to achieving an exponential structure in four spacetime dimensions.
We remark again that we have relied on the four-dimensional Clifford algebra and Dirac spinors throughout our calculation.
In light of this, despite finding an exponential structure in the $(1,i)$ and $(2,1)$ spin structures for general $D$, the second line of \cref{eq:EikExp} should also be understood to hold in the limit $D\rightarrow4$.

\iffalse
Note that there are terms in the square of the leading eikonal that have not been matched to the one-loop result.
These are precisely the terms mentioned in \Cref{sec:Eikonal} that exist outside of the Hilbert space of the one-loop amplitude.
Their interpretation is simple: when calculating the scattering of a spin-1/2 particle, one cannot probe higher than linear order in the spin of that particle.
Hence, terms in the square of the leading eikonal that are quadratic in the spin of either particle have no analog in the one-loop amplitude.
Correcting the square of the leading eikonal, the relation to the one-loop super-classical terms in impact-parameter space is
\begin{align}\label{eq:HsCorrection}
    \lim_{D\rightarrow4}i\chi_{2}&=\lim_{D\rightarrow4}\left[\frac{1}{2}(i\delta_{1})^{2}-\frac{1}{2}i\delta_{1}^{(1,1)}\left(i\delta_{1}^{(1,1)}+2\sum_{j=1}^{3}i\delta_{1}^{(2,j)}\right)-\frac{1}{2}i\delta_{1}^{(1,2)}\left(i\delta_{1}^{(1,2)}+2\sum_{j=1}^{3}i\delta_{1}^{(2,j)}\right)\right.\notag \\
    &\qquad\qquad\left.-\sum_{j,k=1}^{3}\left(1-\frac{1}{2}\delta_{jk}\right)i\delta_{1}^{(2,j)}i\delta_{1}^{(2,k)}\right],
\end{align}
where $\delta_{jk}$ is the Kronecker delta.
As advertised, the Hilbert space correction removes terms of the schematic form $S_{1}^{2},\,S_{2}^{2},\,S_{1}^{2}S_{2},\,S_{1}S_{2}^{2},$ and $S_{1}^{2}S_{2}^{2}$.
Analogous corrections are needed for particles of any spin at loop level.
\fi

\section{Leading eikonal exponentiation from unitarity}\label{sec:Unitarity}

In this section we show that the observed relationship \cref{eq:EikExp} is guaranteed by unitarity at leading order in $\hbar$.
We will see that the prescription for squaring the eikonal in \cref{eq:SquaringEikonal} is eminently compatible with unitarity in momentum space.
We will begin by studying the scattering of two spin-1/2 particles, before extending our analysis to arbitrary spin using massive on-shell variables \cite{Arkani-Hamed:2017jhn,Aoude:2020onz}.

An analysis of the connection between unitarity and the eikonal exponentiation was first presented in the spinless case in ref.~\cite{Cristofoli:2020uzm}.\footnote{I thank Poul Henrik Damgaard for pointing out this initial exploration of the connection between unitarity and the eikonal exponentiation, and for discussions on this connection.}
Ref.~\cite{DiVecchia:2021bdo} later applied similar ideas at the two-loop level to also account for radiation reaction effects.
See also ref.~\cite{Damgaard:2021ipf}.

\subsection{Spin-1/2 $\times$ spin-1/2 scattering}

To begin, we write the $S$ matrix as
\begin{align}
    \cS=1+i\cT.
\end{align}
Requiring that the $S$ matrix be unitary imposes that $\cT$ satisfies $2\,\text{Im}(\cT)=\cT\cT^{\dagger}$.
Noticing that $\cT\sim\mathcal{O}(G)$ for $2\rightarrow2$ scattering, expanding both sides of this condition to $\mathcal{O}(G)$ tells us that the tree-level amplitude must be real.
Going further to $\mathcal{O}(G^{2})$ relates the imaginary part of the one-loop amplitude to the product of two tree-level amplitudes:
\begin{align}
    2\,\text{Im}(\cT_{2})&=\cT_{1}\cT_{1},
\end{align}
where we have used the realness of $\cT_{1}$.

We can convert this to an amplitude by taking the expectation value in external states appropriate for $2\rightarrow2$ scattering of spin-1/2 particles:
\begin{align}
    2(2\pi)^{D}\delta(p_{1}+p_{2}-p_{1}^{\prime}-p_{2}^{\prime})\text{Im}[\cM(q)]&=\langle p_{1}^{\prime},p_{2}^{\prime};s_{1}^{\prime},s_{2}^{\prime}|\cT_{1}\cT_{1}|p_{1},p_{2};s_{1},s_{2}\rangle,
\end{align}
where $q$ is the momentum transfer, $q=p_{1}-p_{1}^{\prime}$.
We have included labels for the polarizations of the external states.
As was mentioned in \Cref{sec:LeadingEikonal}, in order to generate the spinless part of the leading eikonal one needs that the polarizations of the scattered particles are unchanged by the scattering.
Thus we set $s_{1,2}^{\prime}=s_{1,2}$.
The completeness relation for spinors,
\begin{align}
    1&=\int\frac{d^{D-1}k_{1}}{(2\pi)^{D-1}2E_{k_{1}}}\frac{d^{D-1}k_{2}}{(2\pi)^{D-1}2E_{k_{2}}}\sum_{s_{1}^{k},s_{2}^{k}}|k_{1},k_{2};s_{1}^{k},s_{2}^{k}\rangle\langle k_{1},k_{2};s_{1}^{k},s_{2}^{k}|,
\end{align}
where the $k$ superscripts on the polarizations indicate intermediate polarizations,
converts the right-hand side of this expression to a product of amplitudes:
\begin{align}\label{eq:UnitarityCut}
    2\delta(p_{1}+p_{2}-p_{1}^{\prime}-p_{2}^{\prime})\text{Im}&[\cM_{2}(q)]=\int\frac{d^{D-1}k_{1}}{(2\pi)^{D-1}2E_{k_{1}}}\frac{d^{D-1}k_{2}}{(2\pi)^{D-1}2E_{k_{2}}}(2\pi)^{D}\delta(p_{1}+p_{2}-k_{1}-k_{2})\delta(k_{1}+k_{2}-p_{1}^{\prime}-p_{2}^{\prime})\notag \\
    &\quad\times\sum_{s_{1}^{k},s_{2}^{k}}\cM_{1}(k_{1},s_{1}^{k};k_{2},s_{2}^{k}\rightarrow p_{1}^{\prime},s_{1};p_{2}^{\prime},s_{2})\cM_{1}(p_{1},s_{1};p_{2},s_{2}\rightarrow k_{1},s_{1}^{k};k_{2},s_{2}^{k}).
\end{align}
Both amplitudes in the cut are on shell.
We can drop the delta function on the left-hand side, and one of the delta functions on the right-hand side. 
In the former case we understand that the omitted delta function imposes $p_{1}^{\prime}+p_{2}^{\prime}=p_{1}+p_{2}$, and in the latter case it imposes $p_{1}^{\prime}+p_{2}^{\prime}=k_{1}+k_{2}$, which is subsequently fixed to $p_{1}+p_{2}$ by the remaining delta function.
We are left with
\begin{align}\label{eq:UnitarityCutLessDeltas}
    2\,\text{Im}[\cM_{2}(q)]=\int&\frac{d^{D-1}k_{1}}{(2\pi)^{D-1}2E_{k_{1}}}\frac{d^{D-1}k_{2}}{(2\pi)^{D-1}2E_{k_{2}}}(2\pi)^{D}\delta(p_{1}+p_{2}-k_{1}-k_{2})\notag \\
    &\qquad\times\sum_{s_{1}^{k},s_{2}^{k}}\cM_{1}(k_{1},s_{1}^{k};k_{2},s_{2}^{k}\rightarrow p_{1}^{\prime},s_{1};p_{2}^{\prime},s_{2})\cM_{1}(p_{1},s_{1};p_{2},s_{2}\rightarrow k_{1},s_{1}^{k};k_{2},s_{2}^{k}).
\end{align}

The integrand of \cref{eq:UnitarityCutLessDeltas} already looks very similar to the squaring of the leading eikonal in \cref{eq:SquaringEikonal}.
Before we can make the connection, though, we must address the fact that the external states in the cut amplitudes depend on the transfer momenta $q_{1}=p_{1}-k_{1}=k_{2}-p_{2}$ and $q_{2}=k_{1}-p_{1}^{\prime}=p_{2}^{\prime}-k_{2}$, where $q_{1}+q_{2}=q$.
We circumvented this issue in \Cref{sec:LeadingEikonal} by boosting the final-state spinors to have momentum equal to the initial-state spinors, yielding $\mathcal{U}_{i}=1+\mathcal{O}(\hbar)$.
However, one of the amplitudes in the cut in \cref{eq:UnitarityCutLessDeltas} now has initial momenta that depend on the transfer momentum $q_{1}$.
The completeness relation for heavy spinors allows us to relegate this dependence to subleading orders in $\hbar$:
\begin{align}
    \sum_{s_{1}^{k}}u_{v_{1}}(k_{1},s_{1}^{k})\bar{u}_{v_{1}}(k_{1},s_{1}^{k})=\frac{1+\slashed{v}_{1}}{2}+\mathcal{O}(\hbar^{2})=\sum_{s_{1}^{k}}u_{v_{1}}(p_{1},s_{1}^{k})\bar{u}_{v_{1}}(p_{1},s_{1}^{k})+\mathcal{O}(\hbar^{2}),
\end{align}
and analogously for the polarization sum of the other particle.
Boosting now the final-state spinors with momenta $p_{1,2}^{\prime}$ to have momenta $p_{1,2}$, we can describe all on-shell states in \cref{eq:UnitarityCutLessDeltas} with spinors with no dependence on the integration momenta.
Thus we can write
\begin{align}\label{eq:ClassicalUnitarity}
    2\,\text{Im}[\cM_{2}(q)]=\int&\frac{d^{D-1}k_{1}}{(2\pi)^{D-1}2E_{k_{1}}}\frac{d^{D-1}k_{2}}{(2\pi)^{D-1}2E_{k_{2}}}(2\pi)^{D}\delta(p_{1}+p_{2}-k_{1}-k_{2})\notag \\
    &\qquad\times\cM_{1}^{\text{cl.}}(p_{1},s_{1};p_{2},s_{2}\rightarrow k_{1},s_{1}^{k};k_{2},s_{2}^{k})\otimes\cM_{1}^{\text{cl.}}(k_{1},s_{1}^{k};k_{2},s_{2}^{k}\rightarrow p_{1}^{\prime},s_{1};p_{2}^{\prime},s_{2})+\mathcal{O}(\hbar),
\end{align}
where the product $\otimes$ is precisely that in \cref{eq:SquaringEikonal}.
The superscript $\text{cl.}$ indicates that we also truncate the amplitudes to their classical portions and drop ultralocal terms. 

All that remains to obtain \cref{eq:EikExp} is to Fourier transform \cref{eq:ClassicalUnitarity}.
This can be done in an identical fashion to Section 6.2 in ref.~\cite{DiVecchia:2021bdo}, only setting $k=0$ (we work at one-loop order) and hence replacing the five-point amplitudes there simply with our $\cM^{\text{cl.}}_{1}$.
We have included details of this transform -- adapted to our problem -- in \Cref{sec:FTUnitarity}.
Applying \cref{eq:FTUnitarity} to each term in the polarization sum in \cref{eq:ClassicalUnitarity} gives
\begin{align}\label{eq:UnitarityIPSpace}
    \text{Im}[\widetilde{\cM}_{2}(\mathbf{b};s_{1}\rightarrow s_{1};s_{2}\rightarrow s_{2})]&=\frac{1}{2}\delta_{1}(\mathbf{b};s_{1}\rightarrow s_{1}^{k};s_{2}\rightarrow s_{2}^{k})\otimes\delta_{1}(\mathbf{b};s_{1}^{k}\rightarrow s_{1};s_{2}^{k}\rightarrow s_{2})+\mathcal{O}(\hbar).
\end{align}
At leading order in $\hbar$, the left-hand side of \cref{eq:UnitarityIPSpace} is by definition the contribution from the super-classical portion of the one-loop amplitude in impact-parameter space, which we have shown by direct computation in \Cref{sec:LeadingEikonal} to be purely imaginary.
Hence,
\begin{align}
    \text{Im}[\widetilde{\cM}^{\text{sc.}}_{2}(\mathbf{b})]=-i\widetilde{\cM}^{\text{sc.}}_{2}(\mathbf{b})&=\frac{1}{2}\delta_{1}(\mathbf{b};s_{1}\rightarrow s_{1}^{k};s_{2}\rightarrow s_{2}^{k})\otimes\delta_{1}(\mathbf{b};s_{1}^{k}\rightarrow s_{1};s_{2}^{k}\rightarrow s_{2}),
\end{align}
which is precisely \cref{eq:EikExp}.
We have implicitly used the four-dimensional Clifford algebra when truncating the amplitudes in the cut to $\cM^{\text{cl.}}$.
Thus, the result of this analysis is to be thought of in the limit $D\rightarrow4$.

So much for the spin-1/2 $\times$ spin-1/2 case.
Let us extend this analysis to arbitrary spin.

\subsection{Arbitrary-spin scattering}

\Cref{eq:UnitarityCutLessDeltas} already appears rather scalable to arbitrary-spin scattering.
There is only one sticking point: we must be able to express the polarizations of arbitrary-spin states in terms of momenta that are independent of the integration momenta.
We can achieve this explicitly rather easily by making use of the heavy on-shell spinors of ref.~\cite{Aoude:2020onz}.
Given that we will employ the spinor-helicity formalism, our arguments in this subsection are also restricted to the limit $D\rightarrow4$.

As touched on in the introduction, the gravitational Compton amplitude for matter with spin $s\geq 2$ needs additional contact terms to render it well-defined.
However, these contact terms will not affect the factorization properties of the amplitude when a matter propagator is taken on shell.
The unitarity technique we used in the previous subsection, and which we will now apply to higher spins here, cuts the two matter lines.
The arguments made in this subsection for arbitrary spins are therefore not affected by the contact terms needed to fix the spin $s\geq2$ Compton amplitude.

To avoid over-cluttering with notation, let us consider the scattering of a spin-$s$ and a spin-$0$ particle.
Up to ultralocal terms, the tree-level amplitude contributing to the leading eikonal for such a process can be written in terms of three-point amplitudes \cite{Guevara:2018wpp,Guevara:2019fsj,Aoude:2020onz}
\begin{align}\label{eq:FourPointFactorization}
    \cM_{1}^{s\times0}(p_{1,I},p_{2}\rightarrow (p_{1}^{\prime})^{I^{\prime}},p_{2}^{\prime})&=\sum_{h=\pm}\cM^{s}_{\vdash}(p_{1,I}\rightarrow (p_{1}^{\prime})^{I^{\prime}},-q^{h})\frac{i}{q^{2}}\cM^{0}_{\vdash}(p_{2}\rightarrow p_{2}^{\prime},q^{-h}),
\end{align}
where $I$ and $I^{\prime}$ are massive little group indices and $h$ is the helicity of the exchanged graviton.
When referring to a spin $s$ particle, we will understand $I$ and $I^{\prime}$ to represent a set of $2s$ little group indices.
The key point in this case is that the spin-$s$ three-point amplitude can be expressed in terms of the spin-1/2 amplitude in heavy on-shell variables as \cite{Aoude:2020onz}
\begin{subequations}
\begin{align}\label{eq:SpinSThreePoint}
    \cM^{s}_{\vdash}(p_{1,I}\rightarrow (p_{1}^{\prime})^{I^{\prime}},-q^{h})&=(-1)^{2s}\frac{\kappa }{m_{1}}p_{1}^{\mu}p_{1}^{\nu}\varepsilon^{h}_{\mu\nu}\left[\langle 1^{\prime I^{\prime}}_{v_{1}}|^{\alpha}{\left(\hat{\cM}^{1/2}_{\vdash}(-q^{h})\right)_{\alpha}}^{\beta}| 1_{v_{1},I}\rangle_{\beta}\right]^{\odot2s},
\end{align}
where $\odot$ is the symmetrized tensor product \cite{Guevara:2018wpp} and
\begin{align}
    {\left(\hat{\cM}_{\vdash}^{1/2}(-q^{h})\right)_{\alpha}}^{\beta}\equiv{\left(\mathbb{I}-h \frac{q\cdot S_{1}}{m_{1}}\right)_{\alpha}}^{\beta}+\mathcal{O}(\hbar^{2}).
\end{align}
\end{subequations}
The $\mathcal{O}(\hbar^{2})$ corrections arise when the initial residual momentum is not zero, as is the case for the amplitude in the cut with initial momenta $k_{1,2}$.
The spin vector here is defined through the Pauli-Lubanski pseudovector; the matrix element of this spin vector coincides with \cref{eq:SpinDef} \cite{Guevara:2019fsj,Aoude:2020onz}.
We have again normalized the external states to be dimensionless.

The momentum described by heavy on-shell variables is always proportional only to the velocity of the heavy particle \cite{Aoude:2020onz}.
More specifically,
\begin{align}\label{eq:HeavyMomentum}
    m|p_{v}^{I}\rangle_{\alpha}[p_{vI}|_{\dot{\alpha}}=p_{v\alpha\dot{\alpha}},\quad \text{where}\quad p_{v}^{\mu}=\left(1-\frac{k^{2}}{4m^{2}}\right)mv^{\mu}\equiv m_{k}v^{\mu},
\end{align}
where the factor of $m$ on the left hand side of the first relation is just our normalization of the spinors.
A consequence of this is that two heavy spinors describing particles of the same mass and velocity and whose momentum differs only by $k^{\prime}-k\ll m$ satisfy the on-shell condition
\begin{align}
    \langle p^{\prime J}_{v}p_{vK}\rangle=\frac{\sqrt{m_{k}m_{k^{\prime}}}}{m}{\delta^{J}}_{K}={\delta^{J}}_{K}+\mathcal{O}(\hbar).
\end{align}
Therefore, we must fix $I^{\prime}=I$ in order for \cref{eq:SpinSThreePoint} to possess a spin-monopole contribution, analogously to the spin-1/2 case.
We must keep in mind, then, that the raised and lowered $I$ indices are not summed over in the following, as we consider the external polarizations to be fixed.

\Cref{eq:HeavyMomentum} is the key to our goal of expressing the external states in terms of only the initial momentum.
It tells us that the dependence of the heavy external states on infinitesimal momenta is simply encoded in a multiplicative factor, so we can write
\begin{align}\label{eq:BoostHeavySpinors}
    |p_{v}^{I}\rangle=\sqrt{\frac{m_{k}}{m}}|v^{I}\rangle=|v^{I}\rangle+\mathcal{O}(\hbar).
\end{align}
Let us now make the external spinors explicit in the spin-$s$ $\times$ spin-0 integrand analogous to \cref{eq:UnitarityCutLessDeltas}.
This integrand is
\begin{align}
    &\left({\delta^{I_{k}^{\prime}}}_{I_{k}}\right)^{\odot2s}\cM_{1}^{s\times0}(p_{1,I},p_{2}\rightarrow k_{1}^{I_{k}},k_{2})\cM_{1}^{s\times0}(k_{1,I_{k}^{\prime}},k_{2}\rightarrow (p_{1}^{\prime})^{I},p_{2}^{\prime})\notag \\
    &=\left({\delta^{I_{k}^{\prime}}}_{I_{k}}\right)^{\odot2s}\sum_{h_{1},h_{2}}\cM_{\vdash}^{s}(p_{1,I}\rightarrow k_{1}^{I_{k}},-q_{1}^{h_{1}})\frac{i}{q_{1}^{2}}\cM_{\vdash}^{0}(p_{2}\rightarrow k_{2},q_{1}^{-h_{1}})\cM_{\vdash}^{s}(k_{1,I_{k}^{\prime}}\rightarrow (p_{1}^{\prime})^{I},-q_{2}^{h_{2}})\frac{i}{q_{2}^{2}}\cM_{\vdash}^{0}(k_{2}\rightarrow p_{2}^{\prime},q_{2}^{-h_{2}}).
\end{align}
The two graviton momenta satisfy $q_{1}+q_{2}=p_{1}-p_{1}^{\prime}=q$.
The spinor structure is contained in the product of spin-$s$ three-point amplitudes.
Looking closer at this,
\begin{align}
    &\left({\delta^{I_{k}^{\prime}}}_{I_{k}}\right)^{\odot2s}\cM_{\vdash}^{s}(p_{1,I}\rightarrow k_{1}^{I_{k}},-q_{1}^{h_{1}})\cM_{\vdash}^{s}(k_{1,I_{k}^{\prime}}\rightarrow (p_{1}^{\prime})^{I},-q_{2}^{h_{2}})\notag \\
    &\sim\left({\delta^{I_{k}^{\prime}}}_{I_{k}}\right)^{\odot2s}\left[\langle k^{I_{k}}_{v_{1}}|^{\alpha}{\left(\hat{\cM}_{\vdash}^{1/2}(-q_{1}^{h_{1}})\right)_{\alpha}}^{\beta}| 1_{v_{1},I}\rangle_{\beta}\right]^{\odot2s}\left[\langle 1^{\prime I}_{v_{1}}|^{\gamma}{\left(\hat{\cM}_{\vdash}^{1/2}(-q_{2}^{h_{2}})\right)_{\gamma}}^{\delta}| k_{v_{1},I_{k}^{\prime}}\rangle_{\delta}\right]^{\odot2s} \\
    &=\left({\delta^{I_{k}^{\prime}}}_{I_{k}}\right)^{\odot2s}\left[\langle 1^{I_{k}}_{v_{1}}|^{\alpha}{\left(\hat{\cM}_{\vdash}^{1/2}(-q_{1}^{h_{1}})\right)_{\alpha}}^{\beta}| 1_{v_{1},I}\rangle_{\beta}\right]^{\odot2s}\left[\langle 1^{I}_{v_{1}}|^{\gamma}{\left(\hat{\cM}_{\vdash}^{1/2}(-q_{2}^{h_{2}})\right)_{\gamma}}^{\delta}| 1_{v_{1},I_{k}^{\prime}}\rangle_{\delta}\right]^{\odot2s}+\mathcal{O}(\hbar),
\end{align}
where the $\sim$ indicates we're ignoring for now the non-spinor portion of \cref{eq:SpinSThreePoint}.
In the last line we have used \cref{eq:BoostHeavySpinors} to remove the $q_{1,2}$ dependence of the spinors.
The completeness relation for the heavy on-shell spinors $| 1_{v_{1},I_{k}}\rangle_{\delta}\langle 1_{v_{1}}^{I_{k}}|^{\alpha}={\mathbb{I}_{\delta}}^{\alpha}$,\footnote{Recall that we have normalized our spinors to be dimensionless. Otherwise, the right-hand side of this relation would be multiplied by $m_{1}$.} combined with ${\left(q_{1}\cdot S_{1}\right)_{\gamma}}^{\alpha}{\left(q_{2}\cdot S_{1}\right)_{\alpha}}^{\beta}\propto {(q_{1}\cdot q_{2})_{\gamma}}^{\beta}-i{(q_{1\mu}q_{2\nu}\sigma^{\mu\nu})_{\gamma}}^{\beta}=\mathcal{O}(\hbar)$ ensures that we don't have more than one power of the spin between a pair of spinors.

Having expressed all external states independently of the momenta of integration, and truncated to classical, local terms, we can indeed write the analog to \cref{eq:ClassicalUnitarity} for higher-spin scattering.
The remainder proceeds identically to the spin-1/2 case, suggesting the exponentiation of the leading eikonal up to $\mathcal{O}(G^{2})$ for the scattering of an arbitrary-spin particle and a scalar particle.

To summarize, we have seen in this section that the exponential structure of the leading eikonal including spin up to $\mathcal{O}(G^{2})$ is a consequence of the unitarity of the $S$ matrix at leading order in $\hbar$.

\section{Conclusion}\label{sec:Conclusion}

The eikonal phase plays a crucial role in the derivation of observables from scattering amplitudes, including when spin effects are present.
Nevertheless, up until now, the exponentiation properties of the eikonal upon including spin have not been analyzed in a gravitational or high-spin context.
We have initiated a study in this direction, focusing initially on the leading eikonal for spin-1/2 $\times$ spin-1/2 scattering up to one-loop order.
Defining the square of the leading eikonal through \cref{eq:SquaringEikonal}, our results demonstrate that the super-classical amplitude at one-loop order is indeed related to the leading eikonal in a way that suggests exponentiation.
Two of the four classical spin structures satisfy this relation in arbitrary spacetime dimensions.
It is interesting to note that the two that don't are the two which are multiplied by an overall $\mathbf{q}^{2}$, and hence do not enter the leading eikonal independently of other structures.
However, the usage of the four-dimensional Clifford algebra in the manipulation of gamma matrices means that the exponentiation demonstrated herein should be considered in the limit $D\rightarrow4$.
The spin structures proportional to $\mathbf{q}^{2}$ at one-loop satisfy the exponential relation to the leading eikonal only in this limit; the relation is violated in general dimensions by terms that vanish when $D\rightarrow4$.

Further to this direct calculation, we have demonstrated that the suggested exponentiation of the leading eikonal is in fact a consequence of unitarity.
In contrast to the spinless case, in which this relation is more immediate, in the spinning case it is a consequence of the interplay between the completeness relation for spinors and the classical limit.
Handling the classical limit in this context was easily managed by appealing to heavy particle states.
Our prescription for squaring the leading eikonal relies on a polarization sum for finite-spin particles.
An analysis employing the spin-coherent states of ref.~\cite{Aoude:2021oqj} may present a path to generalizing this squaring to the infinite-spin limit, and removing the reliance on polarization sums altogether.

%Both the direct and unitarity calculations demonstrate that, once spin is involved in the scattering, the exponentiation of the leading eikonal must be supplemented by a so-called Hilbert space correction.
%The role of this correction is to subtract terms in higher powers of the leading eikonal that are not present in higher-loop amplitudes.
%While we have only explicitly constructed the correction at $\mathcal{O}(G^{2})$ for spin-1/2 $\times$ spin-1/2 scattering, it's not hard to see that a correction should exist at all orders in the coupling and for arbitrary-spin scattering.
%While having no bearing on phenomenological results, the correction is required to accurately describe the amplitude in impact-parameter space.

Several other extensions to this work come to mind.
First, the arguments in \Cref{sec:Unitarity} appear to be extendable to higher loop orders, which would yield a more direct path to verifying the exponentiation of the leading eikonal, as opposed to computing the leading super-classicalities at each order explicitly.
Such a streamlined approach would be particularly advantageous for the leading eikonal for higher-spin scattering.
Above $\mathcal{O}(G^{2})$ and for spin $s\geq2$ one must take care to properly account for the additional contact terms needed for the Compton amplitude when applying this unitarity technique.
%Relatedly, the Hilbert space correction to higher powers of the leading eikonal is predictable for spin-1/2 $\times$ spin-1/2 scattering at all loop orders.
%It would be interesting to study whether these corrections resum to a closed form, and whether this closed form can be extended to arbitrary spin.

In this note we have focused only on the leading eikonal, though the expression in \cref{eq:AllOrderAmplitude} depends on an exponentiation of contributions at all loop orders.
Studying the exponentiation of subleading eikonal phases with spin is thus crucial to understanding whether \cref{eq:AllOrderAmplitude} -- with the appropriately-modified multiplication of eikonal phases -- remains valid when spin is involved.
Along this line, the combination of the eikonal with spin provides a powerful means for the computation of full (i.e. conservative plus radiative) higher-PM dynamics at low spins \cite{DiVecchia:2021ndb,DiVecchia:2021bdo}.

Yet another direction would investigate the exponentiation of spin structures in general spacetime dimensions, which would require a careful analysis of spin in $D\neq4$.
Such an investigation, along with understanding the exponentiation properties of the sub-leading eikonal, could shed light on why two of the four classical spin structures appear to only exponentiate in $D\rightarrow4$.
A treatment of spin in general dimensions may be sufficient to restore this exponentiation away from $D\rightarrow4$.
Another possibility is that the non-exponentiation in general $D$ is a consequence of the short-range nature of these two spin structures at tree-level, meaning that exponentiation in general $D$ may only be attainable for subleading eikonal phases.

\acknowledgements
I am thankful to Andrea Cristofoli and Poul Henrik Damgaard for enlightening discussions about the eikonal that led to this work, as well as for providing comments on this manuscript.
I also thank Andreas Helset for very helpful discussions and comments on this manuscript.
Finally, I would like to thank Rafael Aoude and Andreas Helset for conversations on related topics.
This project has received funding from the European Union's Horizon 2020 
research and innovation programme under the Marie Sk\l{}odowska-Curie grant 
agreement No. 764850 "SAGEX".

\appendix

\section{Fourier transforms}

In the main text we have made use of Fourier transforms to convert momentum space amplitudes to impact-parameter space.
The rank-0 transform is
\begin{align}
    \int\frac{d^{d}\mathbf{q}}{(2\pi)^{d}}\,e^{i\mathbf{b}\cdot\mathbf{q}}\left(\mathbf{q}^{2}\right)^{\nu}=\frac{2^{2\nu}}{\pi^{d/2}}\frac{\Gamma(\nu+d/2)}{\Gamma(-\nu)}\frac{1}{\left(\mathbf{b}^{2}\right)^{\nu+d/2}}.
\end{align}
The rank-1 and 2 transforms were also necessary for Fourier transforming spin effects.
We find these by differentiating the rank-0 transform with respect to the impact parameter:
\begin{align}
    \int\frac{d^{d}\mathbf{q}}{(2\pi)^{d}}\,e^{i\mathbf{b}\cdot\mathbf{q}}\mathbf{q}^{i}\left(\mathbf{q}^{2}\right)^{\nu}&=-i\frac{\partial}{\partial\mathbf{b}^{i}}\int\frac{d^{d}\mathbf{q}}{(2\pi)^{d}}\,e^{i\mathbf{b}\cdot\mathbf{q}}\left(\mathbf{q}^{2}\right)^{\nu}=i\frac{2^{2\nu+1}}{\pi^{d/2}}\frac{\Gamma(\nu+d/2+1)}{\Gamma(-\nu)}\frac{\mathbf{b}^{i}}{\left(\mathbf{b}^{2}\right)^{\nu+d/2+1}}, \\
    \int\frac{d^{d}\mathbf{q}}{(2\pi)^{d}}\,e^{i\mathbf{b}\cdot\mathbf{q}}\mathbf{q}^{i}\mathbf{q}^{j}\left(\mathbf{q}^{2}\right)^{\nu}&=-\frac{\partial^{2}}{\partial\mathbf{b}^{i}\partial\mathbf{b}^{j}}\int\frac{d^{d}\mathbf{q}}{(2\pi)^{d}}\,e^{i\mathbf{b}\cdot\mathbf{q}}\left(\mathbf{q}^{2}\right)^{\nu}\notag \\
    &=\frac{2^{2\nu+1}}{\pi^{d/2}}\frac{\Gamma(\nu+d/2+1)}{\Gamma(-\nu)}\frac{1}{\left(\mathbf{b}^{2}\right)^{\nu+d/2+1}}\left[\Pi^{ij}-2\left(\nu+1+\frac{d}{2}\right)\frac{\mathbf{b}^{i}\mathbf{b}^{j}}{\mathbf{b}^{2}}\right].
\end{align}
We have introduced the projector
\begin{align}\label{eq:Projector}
    \Pi^{ij}\equiv\delta^{ij}-\frac{\mathbf{p}^{i}\mathbf{p}^{j}}{\mathbf{p}^{2}},
\end{align}
which projects onto the plane orthogonal to the incoming center-of-mass three-momentum.

\section{Loop integrals}\label{sec:LoopIntegrals}

We use the values for $D$-dimensional one-loop integrals in the soft region determined in ref.~\cite{Cristofoli:2020uzm}.
In particular, we have needed the box and cross-box integrals, defined as
\begin{align}
    \mathcal{I}_{\square}&=\int\frac{d^{D}l}{(2\pi)^{D}}\frac{1}{l^{2}(l+q)^{2}[(p_{1}+l)^{2}-m_{1}^{2}][(p_{2}-l)^{2}-m_{2}^{2}]}, \\
    \mathcal{I}_{\boxtimes}&=\int\frac{d^{D}l}{(2\pi)^{D}}\frac{1}{l^{2}(l+q)^{2}[(p_{1}-l-q)^{2}-m_{1}^{2}][(p_{2}-l)^{2}-m_{2}^{2}]}.
\end{align}
Our results depended only on the sum of these two, which is given in ref.~\cite{Cristofoli:2020uzm} to be
\begin{align}
    \mathcal{I}_{\square}+\mathcal{I}_{\boxtimes}&=-\frac{\pi^{1-D/2}}{2^{D+1}m_{1}m_{2}\sqrt{\omega^{2}-1}}\frac{\Gamma(D/2-2)^{2}\Gamma(3-D/2)}{\Gamma(D-4)}\left(-q^{2}\right)^{D/2-3}.
\end{align}
We remark that we have used the mostly-minus metric, in contrast to ref.~\cite{Cristofoli:2020uzm}, where the mostly-plus metric was employed.

\section{Fourier transforming unitarity}\label{sec:FTUnitarity}

For completeness' sake, we show here explicitly the adaptation to our setup of the manipulation employed in Section 6.2 of ref.~\cite{DiVecchia:2021bdo}, taking the unitarity cut in momentum space to a product of eikonal phases in impact-paramter space.
The argument is independent of whether spin is involved in the scattering process, so we omit polarization labels for brevity.
Nevertheless, with an eye to Fourier transforming products of amplitudes with different polarization labels, we do not require the two amplitudes in the cut to be equal.
Our starting point is the convolution of two amplitudes in momentum space by a unitarity cut:
\begin{align}
    \cA(q)=\int\frac{d^{D-1}k_{1}}{(2\pi)^{D-1}2E_{k_{1}}}\frac{d^{D-1}k_{2}}{(2\pi)^{D-1}2E_{k_{2}}}(2\pi)^{D}&\delta(p_{1}+p_{2}-k_{1}-k_{2})\cM(k_{1},k_{2}\rightarrow p_{1}^{\prime},p_{2}^{\prime})\cM^{\prime}(p_{1},p_{2}\rightarrow k_{1},k_{2}).
\end{align}
The momenta carried by the internal graviton legs are $q_{1}=p_{1}-k_{1}=k_{2}-p_{2}$ and $q_{2}=k_{1}-p_{1}^{\prime}=p_{2}^{\prime}-k_{2}$, and are related to the total momentum transfer through $q_{1}+q_{2}=q$.
Both $q_{1,2}$ scale with $\hbar$ in the classical limit.

Now, as was done in ref.~\cite{DiVecchia:2021bdo}, it is convenient to parametrize our momenta in terms of one-dimensional longitudinal components and $(D-2)$-dimensional transverse components, the latter of which are proportional to the transfer momentum $\mathbf{q}$:
\begin{align}
    p_{1}^{\mu}=\left(E_{1},\mathbf{p}^{\perp},p^{L}\right),\quad p_{2}^{\mu}=\left(E_{2},-\mathbf{p}^{\perp},-p^{L}\right),\quad k_{1,2}^{\mu}=(E_{k_{1,2}},\mathbf{k}_{1,2}^{\perp},k_{1,2}^{L}).
\end{align}
The next step is to integrate out the longitudinal components using the delta function.
To do so, we split the delta function in terms of energy, transverse, and longitudinal components:
\begin{align}
    \delta(p_{1}+p_{2}-k_{1}-k_{2})&=\delta(p_{1}^{0}+p_{2}^{0}-k_{1}^{0}-k_{2}^{0})\delta(p_{1}^{\perp}+p_{2}^{\perp}-k_{1}^{\perp}-k_{2}^{\perp})\delta(p_{1}^{L}+p_{2}^{L}-k_{1}^{L}-k_{2}^{L})\notag \\
    &=\delta\left(E_{1}+E_{2}-\sqrt{(\mathbf{k}_{1}^{\perp})^{2}+(k_{1}^{L})^{2}+m_{1}^{2}}-\sqrt{(\mathbf{k}_{2}^{\perp})^{2}+(k_{2}^{L})^{2}+m_{2}^{2}}\right)\delta(\mathbf{k}_{1}^{\perp}+\mathbf{k}_{2}^{\perp})\delta(k_{1}^{L}+k_{2}^{L}).
\end{align}
Integrating over the longitudinal delta function sets $k_{2}^{L}=-k_{1}^{L}$.
To integrate over the energy delta function, we rewrite
\begin{align}\label{eq:EnergyDelta}
    \delta\left(E_{1}+E_{2}-\sqrt{(\mathbf{k}_{1}^{\perp})^{2}+(k_{1}^{L})^{2}+m_{1}^{2}}-\sqrt{(\mathbf{k}_{2}^{\perp})^{2}+(k_{1}^{L})^{2}+m_{2}^{2}}\right)=\frac{E_{k_{1}}E_{k_{2}}}{|k_{1}^{*}(E_{k_{2}}+E_{k_{1}})|}\left[\delta(k_{1}^{L}-k_{1}^{*})+\delta(k_{1}^{L}-k_{2}^{*})\right],
\end{align}
where $k_{1,2}^{*}$ are the roots of the argument of the delta function on the left-hand side, and satisfy $k_{1}^{*}=-k_{2}^{*}$.
By definition of the transverse direction, we have $\mathbf{p}^{\perp}\propto\mathbf{q}$.
Furthermore, since $k_{1,2}$ are related to $p_{1,2}$ by the addition/subtraction of momenta scaling with $\hbar$, we can safely write $|\mathbf{k}_{1}^{\perp}|=|\mathbf{p}^{\perp}-\mathbf{q}_{1}^{\perp}|\ll|\mathbf{p}|$ and $|\mathbf{k}_{2}^{\perp}|=|-\mathbf{p}^{\perp}+\mathbf{q}_{1}^{\perp}|\ll|\mathbf{p}|$, where these inequalities hold component-wise.
Therefore, we find $k_{1}^{*}=-k_{2}^{*}=|\mathbf{p}|+\mathcal{O}(\hbar)$.
Since the transfer momenta obey $|\mathbf{q}_{1,2}|\ll|\mathbf{p}|$ in the classical limit, the solution $k_{2}^{*}$ is outside the domain of the problem, and we can ignore the second delta function on the right-hand side of \cref{eq:EnergyDelta}.
Integrating over the energy delta function thus gives
\begin{align}
    \cA(q)=\int\frac{d^{D-2}k_{1}}{(2\pi)^{D-2}}\frac{d^{D-2}k_{2}}{(2\pi)^{D-2}}\frac{(2\pi)^{D-2}\delta(\mathbf{k}^{\perp}_{1}+\mathbf{k}^{\perp}_{2})}{4\mathbf{p}|E_{k_{2}}+E_{k_{1}}|}\cM(k_{1},k_{2}\rightarrow p_{1}^{\prime},p_{2}^{\prime})\cM^{\prime}(p_{1},p_{2}\rightarrow k_{1},k_{2}).
\end{align}
By energy conservation we must have $E_{k_{1}}+E_{k_{2}}=E$, the total energy of the scattering.
The denominator of the integrand is thus $4\mathbf{p}E=4m_{1}m_{2}\sqrt{\omega^{2}-1}$.

At this point, we can change the variables of integration using $k_{1}=q_{2}+p_{1}^{\prime}$, $k_{2}=q_{1}+p_{2}$, and the fact that $p_{1}^{\prime}$ and $p_{2}$ are constant:
\begin{align}
    \cA(q)=\frac{1}{4m_{1}m_{2}\sqrt{\omega^{2}-1}}\int\frac{d^{D-2}q_{1}}{(2\pi)^{D-2}}\frac{d^{D-2}q_{2}}{(2\pi)^{D-2}}(2\pi)^{D-2}\delta(\mathbf{q}^{\perp}_{1}+\mathbf{q}^{\perp}_{2}-\mathbf{q})\cM(k_{1},k_{2}\rightarrow p_{1}^{\prime},p_{2}^{\prime})\cM^{\prime}(p_{1},p_{2}\rightarrow k_{1},k_{2}).
\end{align}
In the delta function here we have used $\mathbf{p}_{1}^{\prime\perp}=\mathbf{p}^{\perp}-\mathbf{q}$.
Fourier transforming both sides,
\begin{align}\label{eq:FTUnitarity}
    \widetilde{\cA}(\mathbf{b})&=\frac{1}{(4m_{1}m_{2}\sqrt{\omega^{2}-1})^{2}}\int\frac{d^{D-2}q}{(2\pi)^{D-2}}\frac{d^{D-2}q_{1}}{(2\pi)^{D-2}}\frac{d^{D-2}q_{2}}{(2\pi)^{D-2}}e^{i\mathbf{q}\cdot\mathbf{b}}(2\pi)^{D-2}\delta(\mathbf{q}^{\perp}_{1}+\mathbf{q}^{\perp}_{2}-\mathbf{q})\notag \\
    &\qquad\qquad\qquad\qquad\qquad\qquad\times\cM(k_{1},k_{2}\rightarrow p_{1}^{\prime},p_{2}^{\prime})\cM^{\prime}(p_{1},p_{2}\rightarrow k_{1},k_{2})\notag \\
    &=\left[\frac{1}{4m_{1}m_{2}\sqrt{\omega^{2}-1}}\int\frac{d^{D-2}q_{1}}{(2\pi)^{D-2}}e^{i\mathbf{q}_{1}^{\perp}\cdot\mathbf{b}}\cM(p_{1},p_{2}\rightarrow k_{1},k_{2})\right]\notag \\
    &\quad\times\left[\frac{1}{4m_{1}m_{2}\sqrt{\omega^{2}-1}}\int\frac{d^{D-2}q_{2}}{(2\pi)^{D-2}}e^{i\mathbf{q}_{2}^{\perp}\cdot\mathbf{b}}\cM^{\prime}(k_{1},k_{2}\rightarrow p_{1}^{\prime},p_{2}^{\prime})\right]\notag \\
    &=\widetilde{\cM}\widetilde{\cM}^{\prime}.
\end{align}
We have refrained from labelling either side of this equation as an eikonal phase, as one only obtains an eikonal phase upon performing a polarization sum.
We have left both $2\rightarrow2$ amplitudes arbitrary; the crucial point is that the initial and final momenta of each differ by a transfer momentum of $\mathcal{O}(\hbar)$.

This calculation has adapted the approach of ref.~\cite{DiVecchia:2021bdo} to the one-loop computation of relevance to us.
It shows that the convolution of two $2\rightarrow2$ amplitudes in momentum space becomes a product in impact-parameter space.

\bibliography{spineikonal}

\end{document}